\documentclass[aip,reprint]{revtex4-1}
\usepackage[T1]{fontenc}
\usepackage{bm}
\usepackage{graphicx}
\usepackage{amssymb}
\usepackage{leftidx}
\usepackage{upgreek}
\usepackage{siunitx}
\usepackage{amsfonts}
\usepackage{amsmath}
\usepackage[outdir=./]{epstopdf}
\usepackage{color}
\usepackage[utf8]{inputenc}
\usepackage[english]{babel}

\begin{document}


\title{Spatially resolved optical spectroscopy in extreme environment of low temperature, high magnetic fields and high pressure} 



\author{I. Breslavetz}
\author{A. Delhomme}
\author{T. Pelini}
\author{A. Pawbake}
\author{D. Vaclavkova}
\author{M. Orlita}
\author{M. Potemski}
\affiliation{LNCMI, UPR 3228, CNRS, EMFL, Université Grenoble Alpes, 38000 Grenoble, France}
\author{M.-A. Measson}
\email[Authors to whom correspondence should be addressed:
\href{mailto: marie-aude.measson@neel.cnrs.fr}{marie-aude.measson@neel.cnrs.fr} and \href{clement.faugeras@lncmi.cnrs.fr}{clement.faugeras@lncmi.cnrs.fr}]{ }
\affiliation{Université Grenoble Alpes, Institut Neel, CNRS, 38000 Grenoble, France}
\author{C. Faugeras}
\email[Authors to whom correspondence should be addressed:
\href{mailto: marie-aude.measson@neel.cnrs.fr}{marie-aude.measson@neel.cnrs.fr} and \href{clement.faugeras@lncmi.cnrs.fr}{clement.faugeras@lncmi.cnrs.fr}]{ }
\affiliation{LNCMI, UPR 3228, CNRS, EMFL, Université Grenoble Alpes, 38000 Grenoble, France}

%


\date{\today}

\begin{abstract}
We present an experimental set-up developed to perform optical spectroscopy experiments (Raman scattering and photoluminescence measurements) with a micrometer spatial resolution, in an extreme environment of low temperature, high magnetic field and high pressure. This unique experimental setup, to the best of our knowledge, allows us to explore deeply the phase diagram of condensed matter systems by tuning independently these three thermodynamic parameters, while monitoring the low-energy excitations (electronic, phononic or magnetic excitations), to spatially map the Raman scattering response or to investigate objects with low dimensions. We apply this technique to bulk FePS$_3$, a layered antiferromagnet with a N\'{e}el temperature of $T\approx120$~K.
\end{abstract}

\pacs{}

\maketitle 

\section{Introduction}

 Extreme conditions of low temperatures, high magnetic field, high pressure or high doping, in condensed matter physics are sought-after as they allow us to drive electronic systems into exotic electronic ground states that cannot exist/emerge in other conditions. Relevant examples of such ground states include the appearance of superconductivity in metals at cryogenic temperatures or in ultra-doped two dimensional semiconductors~\cite{Costanzo2016}, magnetic orders that build up in magnetic materials below a critical temperature and once established, require high external magnetic fields to be altered~\cite{hagi1999}, pressure induced superconductivity such as, for instance, in Fe-based compounds like FePSe$_3$~\cite{Wang2018}, or the coupling between different electronic ground states such as superconductivity and charge density wave in transition metal dichalcogenides like TaS$_2$~\cite{Grasset2019}. Particular condensed matter systems also exhibit a rich phase diagram with a triple quantum critical endpoint that can only be established in well-defined pressure-temperature-magnetic field conditions~\cite{Mackenzie2005}. Reaching such extreme experimental conditions to perform spatially resolved optical spectroscopy can be a real challenge, and combining them within the same experiment is of course even more challenging.

 In the last fifteen years, microscopic techniques have been implemented in low-temperature and high-magnetic-field environments produced by resistive solenoids, mainly thanks to the development of commercial piezo inertial motors~\cite{Meyer2005} compatible with cryogenic temperature and high magnetic fields, and allowing for sub-micrometer displacements~\cite{berciaud2014,kim2013,Liu2021}. This opened the possibility to perform spatially resolved experiments at low temperatures such as magneto-photoluminescence, -reflectance, -Raman scattering, also with time resolution~\cite{Zhang2017}. High pressures can be obtained using a Diamond Anvil Cell (DAC)~\cite{Jayaraman83}, which is also compatible with optical measurements as diamond is transparent over a broad interval of energy including the visible and infrared ranges. A recent example is the optical investigation at room temperature of a van der Waals heterobilayer in a DAC for pressures up to $\sim 4$~GPa, see Ref.~[\onlinecite{Xia2020}]. Low temperatures for optical measurements with a DAC can be achieved in He4 environments with~\cite{Hudl2015} or without~\cite{Li2020} moderate magnetic fields.

In this manuscript, we present an experimental setup for optical measurements with a micrometer spatial resolution in an extreme environment of low temperature (He4 cryostat), high magnetic fields (compatible with $50$-mm bore resistive magnets at LNCMI-Grenoble producing fields up to $B=31$~T) and high pressure with the use of a DAC. This system allows for spatial mapping of the optical response in such conditions, or to investigate the optical response at a specific location (nanostructures, patterned surface, etc ...) while tuning the extreme conditions.

\section{Experimental setup}

\begin{figure}
 \includegraphics[width=0.9\linewidth,angle=0,clip]{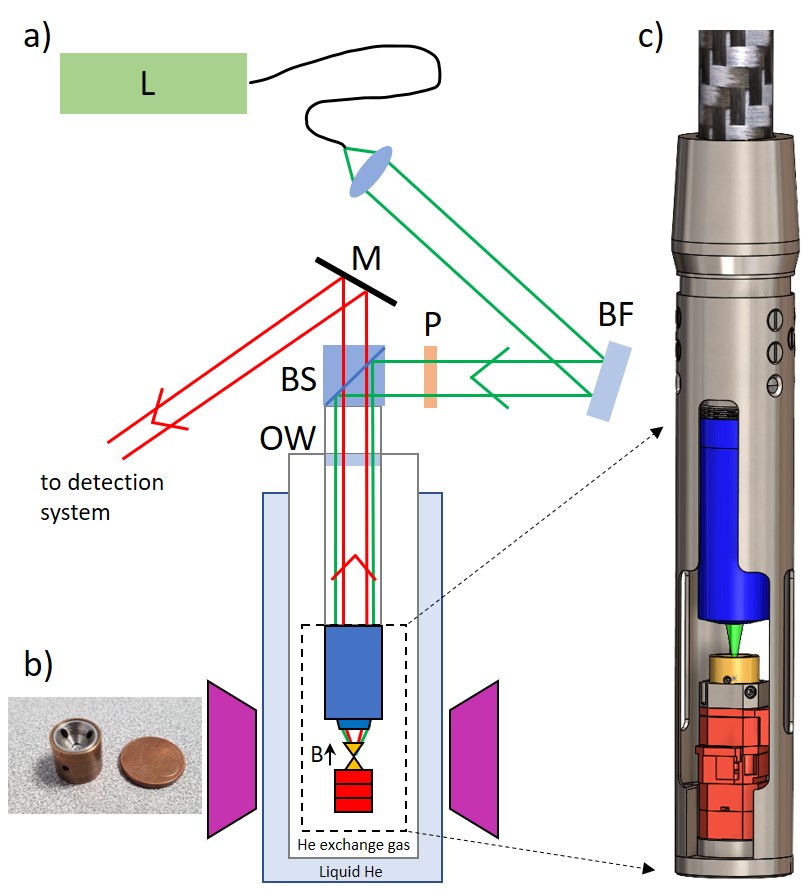}
 \caption{a) Schematics of the experimental setup where L is a laser system, BF a Bragg filter to spectrally clean the excitation laser, P are polarizers or optical filters, OW an optical window, BS a 30/70 (R/T) beam splitter and M a metallic mirror. For reflectance measurements, BF is changed for a metallic mirror and the laser is replaced by a white light source. The magnet is presented in purple. b) Optical photograph of the pressure cell with a 1 Euro cent coin. c) Schematic of the sample holder showing the long working distance objective (blue), the pressure cell (orange) attached to its holder (grey) and the piezo stages (red).}
 \label{Fig_setup}
\end{figure}

Our experimental setup is presented in Fig.~\ref{Fig_setup}a and was conceived to be used in a $50$~mm bore resistive magnet at LNCMI-G. This environment imposes a strong spatial constraint as the field center is located $\sim450$~mm from the top of the magnet, within a bore of $50$~mm. Our setup includes a miniaturized optical table fixed at the top of the experimental probe, and a $1.6$~m tube with an inner diameter of $16$~mm, and hosting the sample holder. We use a monomode optical fiber, at room temperature, to connect the excitation laser source to the top of the probe in order to minimize the effect of vibrations caused by the large flow of deionized water~\cite{Debray2013} (up to $1000$~m$^3$.h$^{-1}$) used to cool down resistive magnets. The excitation beam at the output of the fiber is collimated by an optical lens, cleaned by a reflection on a volume Bragg filter for photoluminescence or Raman scattering experiments, and then, it is sent to the sample holder via a carbon-fiber tube, after a reflection on a $30/70$ beam-splitter. For reflectivity measurement, the volume Bragg filter is replaced by a mirror and the laser source by a white lamp. Polarization optics and optical filters can be used in the excitation or in the collection path.

A circular or linear polarization can be imposed before the last reflection on the beam splitter. The collimated excitation beam is then focused on the pressure cell with a $12$~mm working distance objective of numerical aperture NA=$0.35$. The spot size offers a micrometer spatial resolution. A schematic of this part of the probe is presented in Fig.~\ref{Fig_setup}c. The DAC is of the Tozer design~\cite{graf2011}. It is made of Cu-Be and it weights around $7$~g in total (see Fig.~\ref{Fig_setup}b). This small weight allows us to mount it on piezo stages and to move the DAC below the laser spot with a sub-micrometer spatial resolution. Spatially resolved and hyperspectral images of the optical response of the sample within the DAC can then be reconstructed by spatially scanning the DAC below the laser spot. The optical response is collected with the same objective, goes through the beam splitter and is sent to the detection system using free beam optics. For low-energy (below $100$~cm$^{-1}$) Raman scattering measurements, we use three volume Bragg filters in series before the detection system to reject the stray light. For low temperature experiments, the tube holding the objective-DAC-piezo stages is placed in a closed metallic tube ($32$ mm diameter) filled with $100$ to $200$~mbar of helium exchange gas for an efficient thermal coupling. The tube is then immersed in a bath of liquid helium in a $48$~mm cryostat routinely used in the $50$~mm bore of resistive or superconducting magnets. A local heater/thermometer is inserted below the DAC to allow for tuning temperature up to $T\sim80$~K in the previously described conditions.

\begin{figure}
 \includegraphics[width=0.9\linewidth,angle=0,clip]{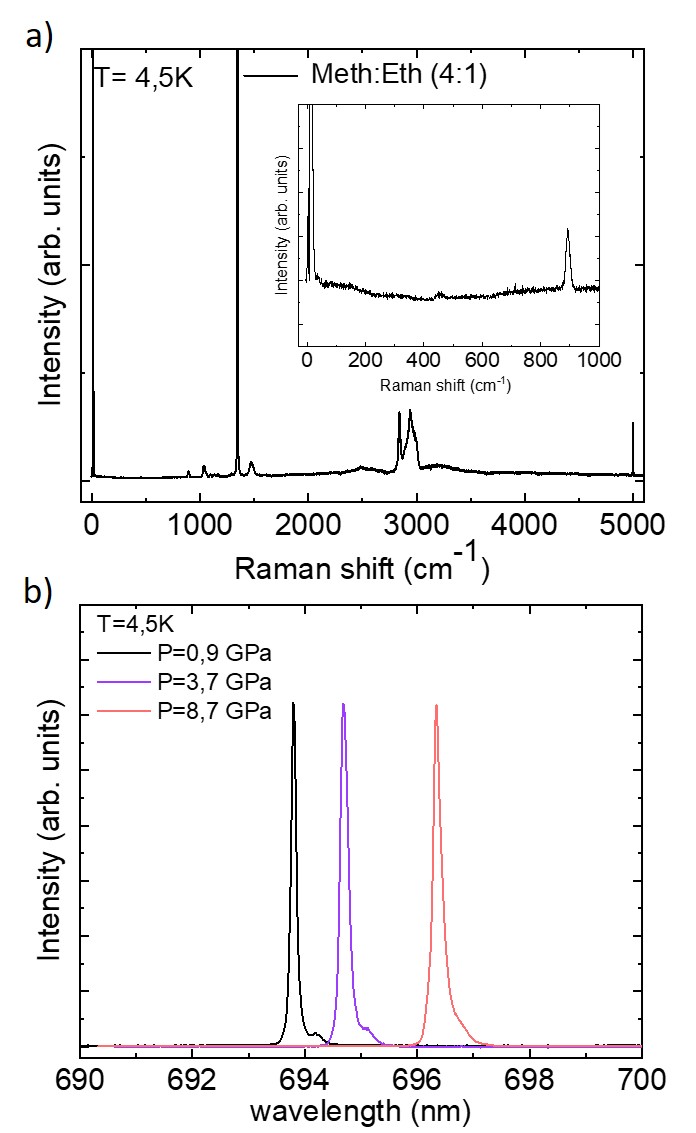}
 \caption{a) Raman scattering spectrum of the pressure transmitting medium (Eth:Meth (1:4)) at $T=4.5$~K and $P=0.9$~GPa. Inset: zoom on the $0 - 800$~cm$^{-1}$ part of the spectrum. b) Photoluminescence spectra of ruby crystals at three different pressures and at $T=4.5$~K.}
 \label{Fig_PTM}
\end{figure}

The gasket has been indented and compressed down to a thickness of $85$~$\mu$m and a hole of $250$~$\mu$m was drilled in its center. The test sample, a triangular flake of bulk FePS$_3$, a layered antiferromagnet, was placed in the hole together with two ruby balls used to measure the pressure~\cite{Syassen2008}. DAC require the use of a pressure transmitting medium (PTM) and we have used a liquid PTM, a mixture of ethanol and methanol (1:4 in volume) expected to remain hydrostatic up to $\sim 10$~GPa, see Ref~[\onlinecite{Klotz2009}]. At $T=4.5$~K, the Raman scattering response of this PTM in the pressure cell is displayed in Fig.~\ref{Fig_PTM}a. The phonon of diamond is clearly visible at $\sim 1340$~cm$^{-1}$, together with the response of the PTM~\cite{Emin2020}. Two energy bands are affected by the response of the PTM, a weak intensity band between $870$ and $1550$ cm$^{-1}$ and a more intense band between $2770$ and $3050$ cm$^{-1}$. In the inset, we highlight the fact that the optical response of this PTM close to the laser line (below $800$~cm$^{-1}$  where most phonon and magnon energies in condensed matter system are found) is nearly featureless, being only composed of a smooth background and allowing for low-energy Raman scattering experiments, see the inset of Fig.~\ref{Fig_PTM}a. These signals are negligible when looking at luminescence signals from direct-band-gap semiconducting materials. The pressure value can be determined by placing the laser spot on the ruby crystal and measuring its photoluminescence signal~\cite{Mao2018}. Fig.~\ref{Fig_PTM}b shows three low temperature spectra of the ruby crystal in the pressure chamber, at selected pressure values.

\begin{figure}
 \includegraphics[width=0.9\linewidth,angle=0,clip]{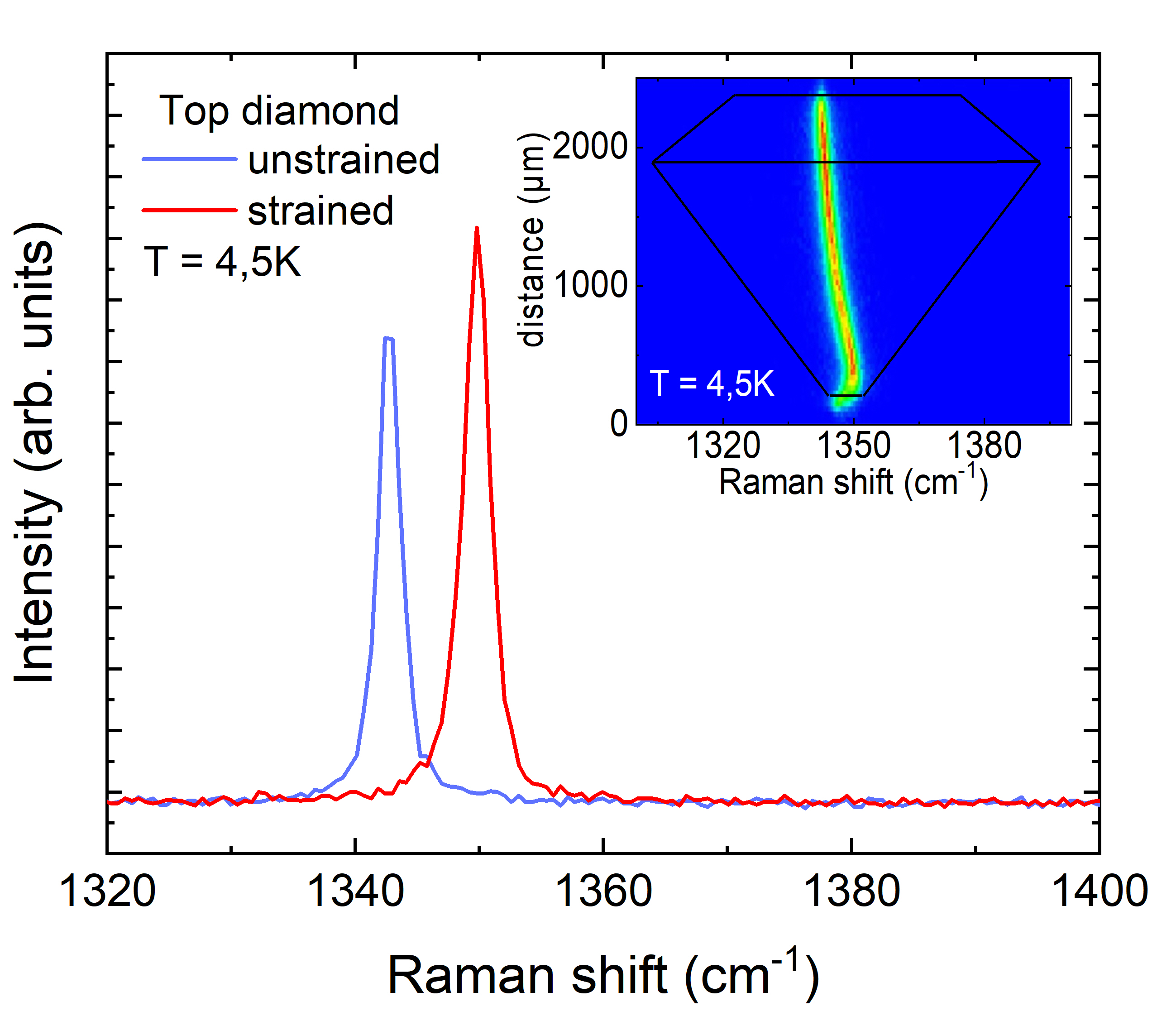}
 \caption{Raman scattering spectra of the top diamond obtained by focusing on the surface (blue spectrum) or close to the pressure chamber (red spectrum) showing the effect of strain on the high energy phonon of diamond. Inset: False color map of the Raman scattering from the high energy phonon in diamond when focusing from the top diamond surface ($z \sim 2000\mu$m) towards the pressure chamber at $z=0$. A schematics of the top diamond is overlaid on the figure.}
 \label{Fig_Diamond}
\end{figure}

Finally, despite the fact that our experimental setup is not confocal, still, the optical response has a depth selectivity which allows us to identify, from the response, if we are focusing on the pressure chamber or not. This can be seen in Fig.~\ref{Fig_Diamond} which displays two characteristic Raman scattering spectra of the top diamond of the DAC. At $T=4.5$~K, depending on the relative distance between the objective and the pressure cell, we can observe changes in the energy position of the doubly degenerate LO-TO phonon at the $\Gamma$ point of diamond, representative of the strain. The strain evolves continuously when focusing deeper into the top diamond, as it is shown in the inset of Fig.~\ref{Fig_Diamond}, and reaches a maximum at the pressure chamber location $\sim 2$~mm below the top surface of the top diamond of the anvil. This effect allows us to identify the correct objective-DAC distance in our experimental setup in which no optical imaging is implemented. After fixing the focus distance, one can spatially map the optical response to locate the ruby crystals in the pressure chamber.

\section{Results}

\begin{figure}
 \includegraphics[width=1\linewidth,angle=0,clip]{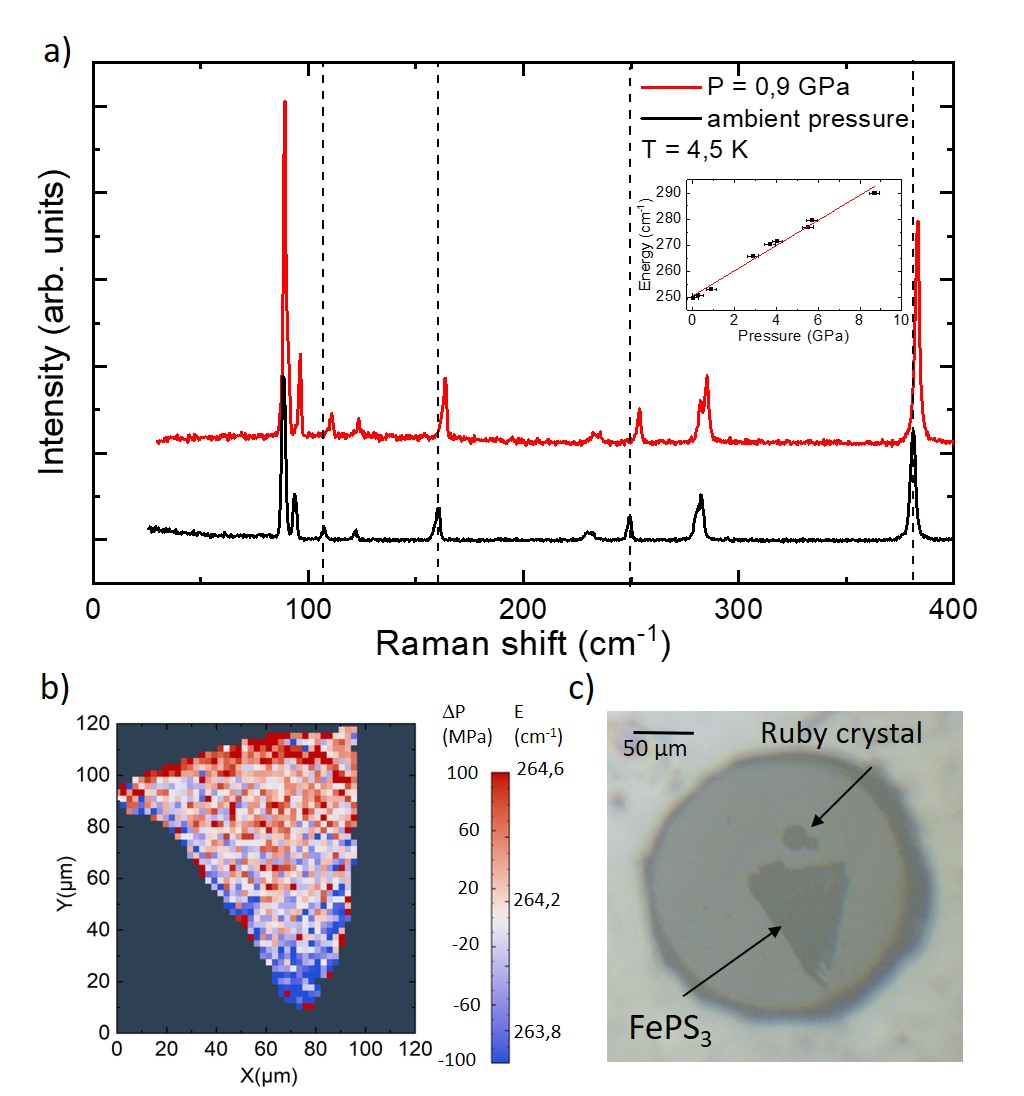}
 \caption{a) Raman scattering spectra of bulk FePS$_3$ measured at $T=4.5$~K and at ambient pressure (black line) and at $P=0.9$~GPa (red line). Vertical dashed lines indicate selected phonon energies without applied pressure. Inset: pressure dependence of the $\sim 260$~cm$^{-1}$ phonon in FePS$_3$, redline is a linear fit of this evolution with a slope of $5.1$cm$^{-1}$/GPa.  b) False-color map of the energy of the $\sim 260$~cm$^{-1}$ phonon in FePS$_3$ at $P=0.9$~GPa. A gradual change of the phonon energy from the center of the pressure chamber to its border clearly indicates a gradient of pressure within the chamber. The color scale is presented as the energy position of this phonon in cm$^{-1}$ or as associated relative pressure change within the chamber in MPa. c) Optical image of the pressure chamber containing the sample (FePS$_3$, of a triangular shape in the image), ruby balls (of a circular shape) to measure the pressure, and loaded with a mixture of ethanol:methanol of 1:4 in volume.}
 \label{Fig_Map1}
\end{figure}

The low-temperature Raman scattering response of FePS$_3$ includes a doubly degenerate antiferromagnetic magnon excitation close to $122$~cm$^{-1}$ identified by both its peculiar temperature dependence~\cite{Sekkin1990}, and recently, by its evolution with an applied magnetic field~\cite{McCreary2020,Liu2021,Diana2021}. The properties of FePS$_3$ under pressure have been described recently from the viewpoints of X-ray diffraction~\cite{Haines2018} and of neutron scattering~\cite{Coak2021}. We present in Fig.~\ref{Fig_Map1}a two low-temperature Raman scattering spectra of FePS$_3$ measured at ambient pressure (black line), and in the DAC with an applied pressure of $0.9$~GPa (red line). The effect of pressure can be seen through the hardening of the phonon modes. Our experimental setup allows for a spatial mapping of the Raman scattering response and this possibility is displayed in Fig.~\ref{Fig_Map1}b with a falsed color spatial map of the energy position of the phonon near $260$~cm$^{-1}$ at $P=0.9$~GPa. One can compare this mapping of the Raman scattering response with the optical photograph of the pressure chamber in Fig.~\ref{Fig_Map1}c. Our Raman scattering data at different pressures up to $8.7$~GPa presented in the inset Fig.~\ref{Fig_Map1}a, indicate that this phonon energy changes with a rate of $5.1$~cm$^{-1}$/GPa. One can clearly see a gradual change of the phonon energy along the flake, and hence of the effective pressure in the chamber. These data indicate that the pressure is larger in the middle of the chamber than at its borders, close to the metallic gasket, and this pressure difference is evaluated to be $\pm 0.1$~GPa from the center of the chamber to its edge.

\begin{figure}
 \includegraphics[width=1\linewidth,angle=0,clip]{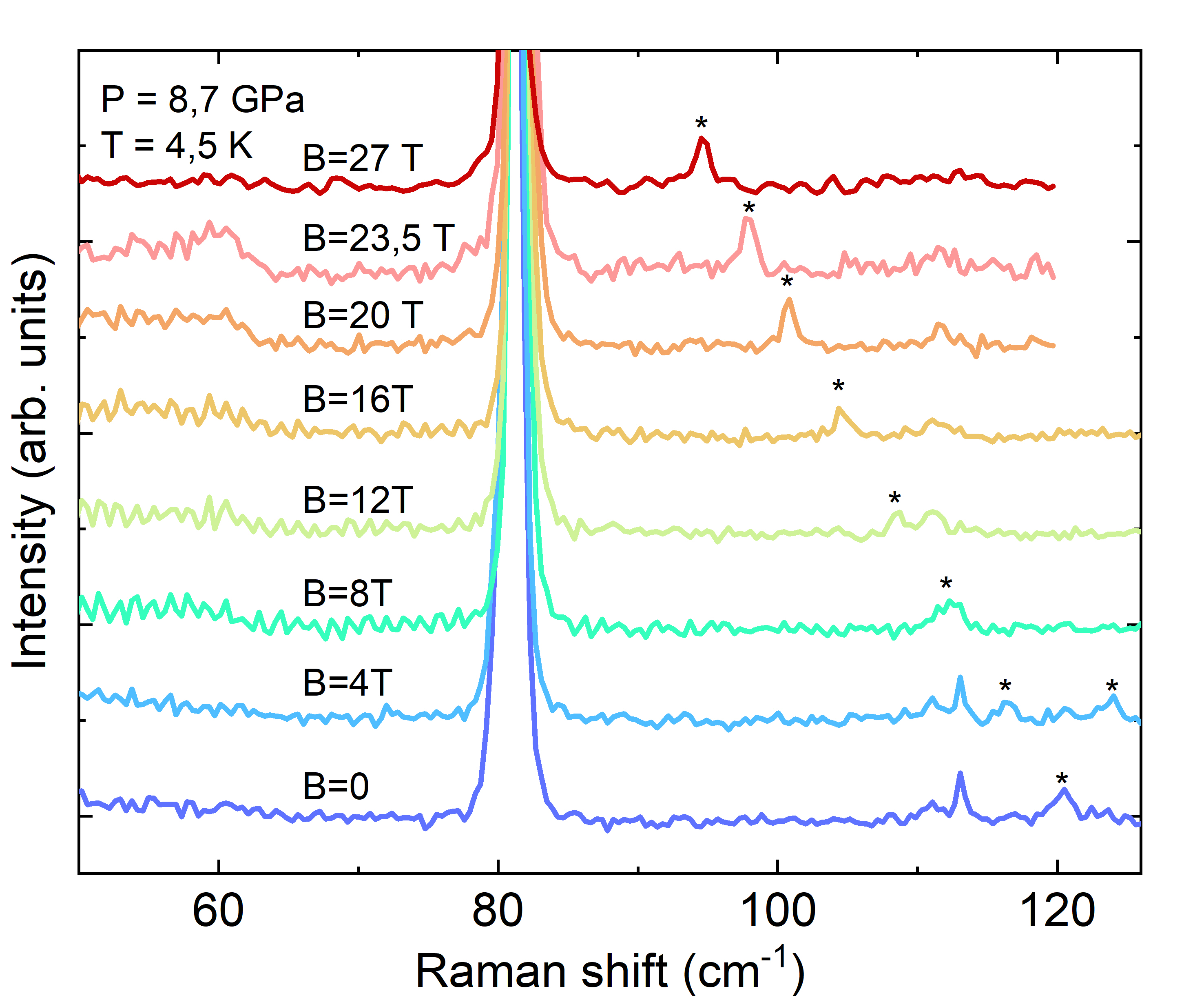}
 \caption{Low temperature ($T=4.5$~K) Raman scattering response of bulk FePS$_3$ at $P=8.7$~GPa for selected values of the magnetic field up to $B=27$~T. Black stars indicate the energy position of the magnetic field dependent magnon excitation.}
 \label{Fig_Bresult}
\end{figure}

When applying a magnetic field transverse to the FePS$_3$ layers, the two-fold degeneracy of the magnon excitation is lifted, the energy of one branch increases with the magnetic field while the energy of the other branch decreases~\cite{McCreary2020}. When a pressure of $8.7$~GPa is set in the DAC, bulk FePS$_3$ is expected to be in the HP1 phase~\cite{Haines2018}. As can be seen in Fig.~\ref{Fig_Bresult}, the magnon excitation is also observed in the Raman scattering response, at an energy very close to that of the magnon at ambient pressure, but within a phonon spectrum modified by the high pressure. When applying a transverse magnetic field the magnon splits into two components and we can observe this excitation up to highest value of the magnetic field. This result illustrates the possibility to probe magnetic and phonon excitations in the very specific environment of low temperature, high pressure and high magnetic fields.

\section{Conclusions}
To conclude, we have presented an experimental set-up for spatially resolved optical investigations (Raman scattering, photoluminescence and reflectance) of condensed matter systems in an extreme environment of low temperature, high magnetic fields provided by the resistive magnets of LNCMI-Grenoble (France), and of high pressure produced in a diamond anvil cell. This set up allows for the spatial mapping of the optical response of the sample with a sub-micrometer resolution at $B=0$ or with an applied magnetic field, or to investigate a well-defined location or systems with reduced dimensions such as van der Waals heterostructures. This work unlocks for condensed matter physics the novel and exciting possibility of inducing exotic electronic ground states at low temperature and high pressure, and to trace the evolution of their elementary excitations as a function of the magnetic field produced by a superconducting or a resisitive magnet, by optical means. Additional functionalities such as measurements of reflectance or photoluminescence with a time resolution are in the process of being implemented.

\section{Acknowledgement}
This work has been partially supported by the EC Graphene Flagship project and by the ANR projects ANR-17-CE24-0030 and ANR-19-CE09-0026. M-A. M. acknowledges the support from the ERC (H2020) (Grant agreement No. 865826). We acknowledge the support of the LNCMI-EMFL, CNRS, Univ. Grenoble Alpes, INSA-T, UPS, Grenoble, France.

\section{Data availability}
The data that support the findings of this study are available from the corresponding author upon reasonable request.

\end{document}